\documentstyle[12pt]{article}
\newcommand{\beq}{\begin{equation}}
\newcommand{\eeq}{\end{equation}}

\newcommand{\half}{{1\over 2}}
\newcommand{\pdr}{\partial}
\newcommand{\Tr}{\;\hbox{\rm  Tr}\;}

\begin{document}
\centerline{\bf  Non--trivial Fixed Point in Four Dimensional Scalar Field
Theory and the Higgs Mass
}

\centerline{ S. G. Rajeev}
\centerline{\it Department of Physics and Astronomy}
\centerline{\it University of Rochester,Rochester,NY 14627-0171}

We show, using the large $N$  limit, that there is a non--trivial scale
invariant action  for four dimensional scalar field theory.
We investigate the possibility  that the scalar sector of the standard model of
particle physics
has such a scale invariant action, with scale invariance being spontaneously
broken by the vacuum expectation value of the scalar. This leads to  a
prediction for the mass of the lightest massive scalar particle (the Higgs
particle) to be $5.4$  Tev.

\centerline{PACS numbers: 14.80.B,12.60.F,03.70}

\vfill\eject

We will study the possibility of a non--trivial ultra--violet stable fixed
point in a four dimensional scalar field theory. We will show, to leading
order in the ${1\over N}$ expansion,\cite{zinn} that such a fixed point does
exist.
A non--trivial fixed point in three dimensional scalar field theory was found
by Wilson and Fisher; it describes second order phase transitions. These
classic studies of renormalization did not find a fixed point in the four
dimensional case\cite{wilson}.
We studied the three dimensional  case by the large $N$ method, constructing
the corresponding three dimensional conformal field theory
in leading order of the ${1\over N}$ expansion\cite{3dconf}.
There are   theorems   that constrain the existence of a  fixed point in  four
dimensional scalar theory \cite{Frohlich}. These analyses are rigorous and
non-perturbative,  and our results do not   contradict them.
  The triviality arguments  apply  to $\lambda\phi^4$ type of interactions. The
self--interactions of the theory we describe are not of this type. The
interaction potential is not a polynomial, or even a real analytic function, of
the field $\phi$. It is  a function of $\phi^2$ ($\phi$ being the scalar field)
with a logarithmic branch point at  $\phi=0$ and grows like $\phi^4\over \log
\phi^2$ for large $\phi^2$.( The form of the potential depends on the
regularization scheme used, and we just described the situation in zeta
function regularization. See below for details.)
Such interactions are not treatable within perturbation theory, indeed the
potential has no power series expansion in the field $\phi^2$;  they  are not
included in the studies that find a negative result on non--trivial fixed
points.

Our result has an application to the standard  model\cite{standard}.
The definition of the standard model by perturbative renormalization  is
`unnatural' \cite{hunters} in the following sense : the values  of the
parameters in the scalar  sector (in paricular the Higgs mass) depend very
sensitively on the short distance (microscopic)
behaviour of the theory.
We can now avoid this difficulty by changing just the scalar self--interaction,
leaving all other aspects of the standard model intact.
Recall that this part of the standard model has not yet been
tested\cite{hunters}.

 The only place in the standard model Lagrangian where scale invariance is
explicitly broken is in the scalar self--interaction. Perturbative
renormalizability requires the scalar self-interaction   to be a quartic
polynomial. We propose that the standard model be modified by choosing  the
scalar self--interaction such that the Lagrangian is scale invariant even after
including quantum corrections. This can be viewed as  a non--perturbative
renormalizability condition.  This fixes the potential, but the answer is not a
polynomial or even an analytic function of the field.
 Scale invariance would then be spontaneously broken by the vacuum expectation
value of the scalar field, just like the internal symmetry. Then it would be
possible to determine the masses of the scalar particles as multiples of this
vacuum expectation value. We will determine
this fixed point potential, to leading order in the ${1\over N}$ expansion.
In a further  approximation,  we will calculate the mass of the lightest
massive scalar.

Let us return to the study of  scalar field theory.
We will consider  a real scalar field $\phi_i$ with $N+1$ components,
$i=0,1,2,\cdots N$ and analyze it in the ${1\over N}$ expansion.
The Lagrangian will be required to  have a global symmetry under the group
$O(N+1)$. Upto some complications of regularization (see below),
\begin{equation}
	L_1=\half \bigg[|\pdr \phi_i|^2+NW\big({\phi^2\over N}\big)\bigg].
\end{equation}
The factors of $N$ have been put in for later convenience.
The self--interactions of the scalar field are contained in the potential $W$;
it is usually taken to be a quadratic in $\phi^2$. We will instead determine it
by the requirement that the Lagrangian be scale invariant even after including
the effects of fluctuations in the field $\phi_i$.

This Lagrangian can be viewed as describing the scalar sector of the standard
model, in the limit where all gauge and Yukawa interactions are ignored. The
scalar field then has four real components, and the global symmetry of the
scalar sector is $O(4)$.

It will be convenient to consider an equivalent Lagrangian with an auxilliary
field $\sigma$,
\begin{equation}
	L_2=\half \big[|\pdr\phi_i|^2+N[\sigma{\phi_i^2\over N}-V(\sigma)]\big]
\end{equation}
where $W$ and $V$ are related by the one--dimensional integral:
\begin{equation}
	\int_0^\infty e^{-N[\sigma \eta-V(\sigma)]}d\sigma=e^{-NW(\eta)}
\end{equation}
where $\eta={\phi^2\over N}$.
In the large $N$ limit $V$ and $W$ will be Legendre transforms of each other.
We will now integrate over all except one of the fields $\phi_i$ leaving
$\sigma$ and $\phi_0=\surd N b$. The resulting effective action will be
\begin{equation}
	S[b,\sigma]=-\ln\int D[\phi]e^{-\int L_2[\phi,b,\sigma] d^4x}
\end{equation}
which is
\begin{equation}
	S[b,\sigma]=N\half \int \big[|\nabla b|^2 d^4x+ \sigma b^2
-V(\sigma)\big]d^4x+\half N\Tr\ln[-\nabla^2+\sigma].
\end{equation}
Here, we have to confront issue of the definition of the divergent quantity
$\Tr\ln[-\nabla^2+\sigma]$. We will define it by the zeta function method
which is technically simple and elegant. It is possible to use  other
regularizations such as momentum cutoff; the details tend to be more
complicated however.

The idea of the zeta function method is that, for $\sigma>0$, the quantity
\begin{equation}
	\zeta(s,x)=<x|[-\nabla^2+\sigma]^{-s}|x>
\end{equation}
is a regular analytic function of $s$ for large enough Re$\; s$. It therefore
defines an analytic function on the complex $s$--plane, which in our case has
only simple pole singularities, located at $s=1,2$. In particular, this
function is regular analytic at $s=0$, so that we can define ($'$ denotes
differentiation with respect to $s$)
\begin{equation}
	\Tr\ln[-\nabla^2+\sigma]=-\int \zeta'(0,x) d^4 x.
\end{equation}
For finite dimensional operators this agrees with the usual definition of the
$\Tr\ln$.

Although this gives a definition of the $\Tr\ln$  without any apparent scale
(such as a momentum cutoff), the result is  in fact {\it not} scale invariant:
there is an anomaly. We can see this most easily by
calculating it for a constant background $\sigma=m^2$:
\begin{equation}
 \zeta(s)=\int {d^4p\over (2\pi)^4}{1\over [p^2+m^2]^s}={1\over 2}{2\pi^2\over
(2\pi)^4}{m^{4-2s}\over (s-2)(s-1)}
\end{equation}
so that by our definition,
\begin{equation}
\int {d^4p\over (2\pi)^4}\ln[p^2+m^2]={1\over 4}{2\pi^2\over
(2\pi)^4}m^4\ln[e^{-3/2}m^2].
\end{equation}
Clearly this is not scale invariant, due to the logarithmic  term.

Now we propose that $V(\sigma)$ be chosen  so as to cancel the vacuum energy of
quantum fluctuations $\Tr\ln[-\nabla^2 +\sigma]$ around a constant  background
$\sigma$.
To leading order in ${1\over N}$,
\begin{equation}
	V(\sigma)={1\over 32\pi^2}\sigma^2\ln[e^{-3/2}\sigma]
\end{equation}
Then the  action $S[b,\sigma]$ is scale invariant under the transformations
\begin{equation}
	b\to \lambda b\quad \sigma\to \lambda^2 \sigma\quad x\to \lambda^{-1}x.
\end{equation}
This can be verified by computing the scale anomaly of
$\Tr\ln[-\nabla^2+\sigma]$  by heat equation methods
and noting that it is cancelled by the variation of the potential. We will give
detailed proofs in a longer paper.

  In the exact theory, $V(\sigma)$ will still be determined by the condition of
cancellation of the energy of constant backgrounds, but it will have
corrections of order ${1\over N}$. It will still be scale invariant, each
contribution to the scale anomaly will be cancelled by a variation of a term in
$V(\sigma)$. This nontrivial fixed point of the renormalization group can thus
be constructed order by order in the ${1\over N}$ expansion.

It is possible to generalize the action to curved spaces preserving  conformal
invariance.
$S[b,\sigma]$  defines a  conformal field theory, although with a nonlocal
action. The three dimensional analogue of this was studied in \cite{3dconf}.
There is however, no trace anomaly in three dimensions, and  there are no
logarithimc terms  in the potential $V(\sigma)$.

The form of the potential that describes the fixed point  depends on the
regularization scheme. Let us consider a momentum cut--off scheme as well, to
make comparison with other approaches easier. If we choose a cut--off function
$K$ on the positive real line( which is one in a neighborhood of  the origin
and zero in a neighborhood of  infinity), we get
\begin{equation}
	V_K(\sigma)={1\over 8\pi^2}\int_0^\infty  \ln[p^2+\sigma]K\big({|p|\over
\Lambda}\big)p^3dp.
\end{equation}
The difference $V_{K_1}-V_{K_2}$ between two choices $K_1$ and $K_2$ will be an
analytic function of $\sigma$ in the neighborhood of $\sigma=0$. This can be
seen by using the Lebesgue dominated convergence theorem, since $K_1-K_2$
is  non--zero  only in some interval $0<a_1<{|p|\over \Lambda}<a_2$. Thus the
`analytic germ' of $V_K$ ( i.e., its equivalence class under the addition of a
function analytic at $\sigma=0$) is independent of the choice of the cut--off
function $K$. If we choose the step function, $K({|p|\over
\Lambda})=\theta(0\leq |p|<\Lambda)$ we get (putting $\Lambda=1$  for
simplicity),
\begin{equation}
	V_K[\sigma]={1\over 32\pi^2}\bigg[\sigma^2\ln[{\sigma\over \surd
e}]+2\sigma\bigg]+ O({\sigma^3})
\end{equation}
This has the same analytic germ as the answer we got in zeta function
regularization.

We can determine the potential $W$ once we know $V$: to leading order it is
just the Legendre transform
\begin{equation}
	W(\eta)=\max_{\sigma}[\eta\sigma-V(\sigma)]
\end{equation}
of $V$. The form of $W$ also depends on the regularization scheme. Consider
first the form of the potential for zeta function regularization.
Then the  extremum occurs at
\begin{equation}
	\eta={1\over 16\pi^2}\sigma\ln\big[{\sigma\over e}\big]
\end{equation}
 which  can be solved recursively for $\eta>0$:
\begin{equation}
\sigma(\eta)={16\pi^2\eta\over \ln\big[{\sigma(\eta)\over e}\big]}
\end{equation}
Once $\sigma$ is determined as a function of $\eta$ this way,
\begin{equation}
	W[\eta]=\half\sigma(\eta)\big[{\sigma(\eta)\over 32\pi^2}+\eta].
\end{equation}
This is not an analytic funtion of $\eta$ at $\eta=0$: the function
$\sigma(\eta)$  has a logarithmic branchpoint at $\eta=0$. For large $\eta$, we
see that
\begin{equation}
	W[\eta]\sim 8\pi^2 {\eta^2\over \log \eta}.
\end{equation}

The non--analytic behaviour of $W[\eta]$   is present in other regularization
schemes as well; the position of  the logarithmic singularity however depends
on the scheme used. If we take as an example
\begin{equation}
	V(\sigma)={1\over 32\pi^2}  \sigma^2\ln{\sigma\over  c_1\surd e}
+c_2\sigma
\end{equation}
the logarithmic branch point is at $\eta=c_2$. For $\eta<c_2$, the function
$W[\eta]$ is multivalued, there are two extrema:if we follow  the smaller of
the two branches, it will vanish identically for  $\eta<\eta_1=c_2-{c_1\over
16\pi^2 e}$.  Thus the kind of potentials we  consider cannot be studied  in an
expansion  in powers of the field variable $\eta={\phi^2\over N}$;
the interaction will look trivial to all orders of perturbation theory in the
momentum cutoff regularization, for example. From now on we will only use zeta
function regularization.

Although the action $S$ is scale invariant, it is possible to have extrema that
break scale invariance.
With the above choice of potential, for constant backgrounds, the problem of
extremizing $S$ reduces to that of extemizing
\begin{equation}
	\half m^2 b_0^2
\end{equation}
where $\sigma=m^2$.
There are two possible solutions: $b_0=0$ and $m$ arbitrary or $m=0$ and $b_0$
arbitrary. In the first case, the $O(N+1)$ symmetry is unbroken and in the
second it can be  spontaneously broken. Let us now study  the spectrum of the
oscillations around this constant background.  This is a subtle problem due to
various Infrared divergences; we will only attempt to get an estimate of the
mass of the lightest massive paricle in the scalar sector.

The  Weyl calculus\cite{folland} can be used to obtain
an expansion for the $\Tr\ln$ in powers of the derivatives of $\sigma$. It will
be convenient to put $\sigma=\chi^2$.
We will study only the longest wavelength fluctuations in $\sigma$, so we will
only calculate the leading term in the expansion of
$\Tr\ln[-\nabla^2+\sigma]-\int V(\sigma)dx$ in terms of the derivatives of
$\sigma$.
We will see that this  term has a simple form  in terms of derivatives of
$\chi$, so that $\chi$ rather than $\sigma$ is the natural variable to use. We
will give details of this rather long calculation in another  paper, noting
here just that the leading order terms in the expansion for the zeta function
are,
\begin{equation}
	\zeta(s)={1\over 16\pi^2}
[{\sigma^{2-s}\over (s-1)(s-2)}-
{1\over 6}\sigma_{ii}\sigma^{-s}+
{1\over 12}s\sigma^{-s}{\sigma_i^2\over \sigma}
]+\cdots .
\end{equation}
Now we have the expansion for the $\Tr\ln$,
\begin{equation}
\Tr\ln[-\nabla^2+\sigma]=
{1\over 32\pi^2}\int \bigg[\sigma^2\ln(e^{-{3\over 2}}\sigma)
+{1\over 6}{\sigma_i^2\over \sigma}\bigg]dx+\cdots .
\end{equation}
Now put $\sigma=\chi^2$ to get
\begin{equation}
\Tr\ln[-\nabla^2+\chi^2]={1\over 32\pi^2}\int \bigg[\chi^4\ln(e^{-{3\over
2}}\chi^2)
+{2\over 3}(\nabla\chi)^2\bigg]dx+\cdots
\end{equation}
upto higher derivatives of $\chi$.

We can combine this with  our original expression for the action,to get,
\begin{equation}
	S[b,\chi]=\half\int[|\nabla b|^2+\chi^2b^2]dx+\half{2\over 3}{1\over
32\pi^2}\int |\nabla\chi|^2 dx+\cdots
\end{equation}
The term with no derivatives in $\chi$ is cancelled by our choice of $V$.
Therefore,  for any constant $b_0$, $b=b_0,\chi=0$ is an extremum. The
oscillations of $\chi$ will then have a mass
\begin{equation}
	m_{\chi}=4\pi\surd 3  b_0\sim 21.7 b_0.
\end{equation}
There could be other modes of oscillation (`particles')  as well, but the
approximation we use should apply only for the lightest massive particle.

Now we consider the possibility that the scale invariant potential $W$
determined as above is  the scalar self--interaction of the standard model.
The standard model would then be `natural', there being an ultraviolet stable
fixed point in the scalar interactions. Since the only source of a scale in the
scalar sector is then the vacuum expectation value, we can predict in principle
the spectrum of scalar particles. This is a difficult problem,
comparable to that of determining the spectrum of Quantum Chromodynamics. It is
a problem  for which  lattice studies of the sort already carried\cite{kuti}
out for $\lambda\phi^4$ would be worthwhile. Here, we make only an estimate of
the mass of the lightest particle in the  scalar sector.
This mass thus describes   the longest wavelength oscillations in the magnitude
of the scalar field. Although there may be several particles (and resonances)
in the scalar sector, this lightest massive particle is the closest thing to a
Higgs particle we would have in our scale invariant theory.
With $b_0\sim 250$ GeV (known from gauge boson masses) we get  for this mass:
\begin{equation}
m_{\chi}\sim 5.4\;\hbox{\rm Tev}.
\end{equation}
An interesting consequence of our analysis is that there is no coupling of the
form $\chi b^2$ in the action. The reflection symmetry $\chi\to -\chi$ is not
spontaneously broken, so there will be no decay of this particle $\chi$
 within the scalar sector. Thus within our approximations this particle is
stable.

We emphasize that since ours is a non--perturbative approach to  the scalar
self--interactions, the usual perturbative unitarity bound\cite{hunters} does
not apply.  There have been several other ideas on fixed points involving
scalars \cite{hill},\cite{nambu}. Our approach only affects the scalar
self--interactions and does not introduce any new parameter in  the standard
model.
We will address generalizations and higher order computations  of this theory
in future publications.

We thank Prof. S. Okubo for discussions,  Prof. L. Orr for suggesting the
Ref.\cite{hunters} and Prof J. Schechter for reading an initial version of this
manuscript.


\begin{thebibliography}{99}

\bibitem{zinn} J. Zinn--Justin,{\it Quantum Field Theory and Critical
Phenomena}, (Clarendon Press, 1996); Nucl. Phys. {\bf B367},105(1991).



\bibitem{wilson} K. G. Wilson, Phys. Rev. {\bf 4}, 3184 (1971); K. G. Wilson
and J. Kogut, Phys. Rep. {\bf 12} 75 (1974).

\bibitem{3dconf} S. Guruswamy, S. G. Rajeev, P. Vitale, Nucl. Phys. {\bf B438},
491 (1995).

\bibitem{Frohlich} R. Fernandez, J. Frohlich and    A. D. Sokal, {\it Random
Walks,Critical Phenomena, and Triviality in Quantum Field Theory}
(Springer--Verlag, 1992).

\bibitem{standard}S. Glashow, Nucl. Phys. {\bf 22}, 579 (1961);A. Salam, in
{\it Proceedings of the 8th Nobel Symposium}(Stockholm) ed. by N. Svartholm
(Almqvist and Wiksell, Stockhom, 1968) p. 367; S. Weinberg, Phys. Rev. Lett.
{\bf 19},1264(1967).

:\bibitem{hunters} J. F. Gunion,H. E. Haber, G. Kane, S. Dawson {\it The Higgs
Huneter's Guide} (Addison-Wesley, New York, 1990).This book contains many
additional references.



\bibitem{folland} G. B. Folland, {\it Harmonic Analysis in Phase Space},
(Princeton, 1989).

\bibitem{kuti} J. Kuti, L. Lin and Y. Shen, Phys. Rev. Lett. {\bf 61} 678
(1988).

\bibitem{hill} W. A. Bardeen and C. T. Hill, Phys. Rev. {\bf D49}, 1437 (1993).

\bibitem{nambu} G. Jona--Lasinio and Y. Nambu Phys. Rev {\bf 124}, 246 (1961);
Y. Nambu in {\it Nagoya 1990, Strong Coupling Gauge Theories and Beyond}, ed.
T. Muta and K. Yamawaki (World Scientific, Singapore, 1991).

\end{thebibliography}
\end{document}